\documentclass[pre,amsmath,amssymb,twocolumn,superscriptaddress]{revtex4}
\usepackage{amsmath,amssymb,graphicx,bm}
\usepackage[usenames]{color}

\newcommand{\Xiamen}{Department of Physics and Key Laboratory of Low Dimensional
Condensed Matter Physics (Department of Education of Fujian Province), Xiamen
University, Xiamen 361005, Fujian, China}
\newcommand{\Lanzhou}{Lanzhou Center for Theoretical Physics, Lanzhou University,
Lanzhou 730000, Gansu, China}

\begin{document}

\title{Thermal rectification in the one-dimensional nonlinearly graded
rotor lattice robust in the thermodynamical limit}
\author{Zhengang Lu}
\affiliation{\Xiamen}
\author{Jiao Wang}
\email{phywangj@xmu.edu.cn}
\affiliation{\Xiamen}
\affiliation{\Lanzhou}

\begin{abstract}
Recently, it has been shown that in graded systems, thermal rectification (TR) effect may
remain in the thermodynamical limit. Here, by taking the one-dimensional rotor lattice as
an illustrating model, we investigate how the graded structure may affect the TR efficiency.
In particular, we consider the case where the interaction is assigned with nonlinear polynomial
functions. It is found that TR is robust in the thermodynamical limit and meanwhile its
efficiency may considerably depend on the details of the graded structure. This finding
suggests that it is possible to enhance the TR effect by taking into account the nonlinear
graded structure even in large systems.
\end{abstract}

\maketitle

\section{Introduction}

Thermal rectification (TR) is an interesting heat transport phenomenon
~\cite{Starr1936, Peyrard2002}. In a system where TR takes place -- such a system is termed
as a thermal rectifier or a thermal diode -- the heat current flows preferably along a
particular direction than along the others. As such TR can be utilized to control and manage
thermal flows, hopefully leading to the promising novel and exciting applications~\cite{Modrev}.

The study of TR is also of fundamental theoretical interest in revealing the basic transport
properties. In this respect, since the pioneer work by Terraneo {\it et al}~\cite{Peyrard2002}
attempting to relate TR with the underlying microscopic dynamics, significant progress has
been made. So far two necessary conditions for TR have been identified~\cite{PeyrardEPL}:
One is the asymmetry in the system's structure and another is the (sensitive) dependence
of the local heat transport on the local structure and temperature. Based on this understanding,
most of the ensuing research has devoted to identifying the various mechanisms that can magnify
the structure asymmetry, e.g., by introducing the long range interactions~\cite{SD2015}, or
the dependence sensitivity of the local heat transport on the local structure and temperature,
e.g., by taking advantage of the phase transition~\cite{Kobayashi}. The study along this
line has turned out very successful and fruitful.

In spite of the progress achieved, however, there is an unsolved theoretical mystery. That
is, for a lattice system, why, as the system size increases, does the TR effect usually decay 
and vanishe eventually in the thermodynamical limit~\cite{ZhYPRL}? Intuitively, provided other
conditions unchanged, when the system size is increased, the change rate of any relevant
quantity along the system would decrease correspondingly, so that the effect of the structural
asymmetry as well as the sensitive dependence of the local heat transport could be weakened.
Given this, in order to retrieve TR in large systems, its two preconditions must be robust
against the increase of the system size. From this consideration, two particular approaches
have been proposed for maintaining TR in large systems. One is to employ the integrability,
because for an integrable system, the heat current flowing across it does not depend on the
system size~\cite{Lepri2003}. An illustrating example is given in Ref.~\cite{SD2018}, where
it is shown that by introducing a harmonic (integrable) chain as a spacer into a thermal diode,
TR keeps its efficiency in any long system as the length of the spacer can be set at will.
Another interesting example consists of hard-point particles with graded masses, which is
not exactly integrable but tends asymptotically to the integrable limit as the system size
increases, where TR is found to hold its efficiency as well~\cite{WJ2012}. The second approach
is to take good advantage of the sensitive temperature dependence of the heat conduction. The
one-dimensional rotor lattice~\cite{RL85, RL87} serves as an enlightening example, where the
sensitive temperature dependence could even be progressively enhanced in the thermodynamical
limit~\cite{RL-Livi, RL-Savin, RL-Spohn, RL-Dhar}. As a result, it is possible to design a
graded rotor lattice whose TR efficiency could even keep increasing instead as the system
size~\cite{SJ2020}.

Note that this intriguing TR effect exhibited in the graded rotor lattice is due to the
strong nonlinear effect of a transition~\cite{RL-Livi, RL-Savin, RL-Spohn, RL-Dhar}. Even
for a linearly graded lattice whose gradient decreases with the increasing system size, the
strong nonlinear effect may manifest itself remarkably (e.g., as seen in the temperature
profile~\cite{SJ2020}). It is therefore interesting to investigate how robust and sensitive
is the underlying mechanism by which the nonlinear effect plays its role to result in TR. 
To probe it, we may turn to the executable question that is concerned with: If the linearly 
graded structure is perturbed, how the TR effect may respond. In previous studies, the 
linearly graded structure has been taken into account intensively and prevailingly, 
which might be out of the implicit assumption that the effect caused by the perturbation 
is negligible or trivial. Our motivation here is to study this issue by considering the 
perturbed graded rotor lattice, and as shown in the following, it is found that this 
assumption does not apply.

\begin{figure}[!]
\includegraphics[width=8.5cm]{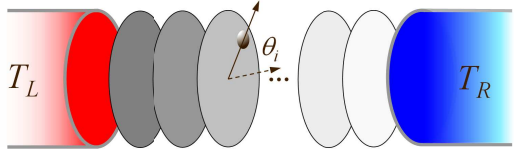}
\caption{The schematic plot of the interaction graded rotor lattice coupled with two heat
baths. For visualization purposes, a rotor is represented by a mass point fixed on a rigid,
massless disk rotating around the horizontal axis, and the interaction between two neighboring
rotors is assumed to be stronger (weaker) if the associated disks are darker (lighter).}
\end{figure}

\section{Model and method}

Our model system consists of $N$ rotors positioned on a one-dimensional lattice (see Fig.~1
for a schematic plot). We take the dimensionless units throughout, in which the lattice
constant is unit and thus $N$ also measures the length (size) of the system. The Hamiltonian
for a symmetric, homogeneous rotor lattice is~\cite{RL85, RL87}
\begin{equation}
H=\sum_i [\frac{1}{2}\dot{\theta}^2_i+V(\theta_{i+1}-\theta_i)],
\end{equation}
where $\theta_i$ is the angular variable of the $i$th rotor with respect to a given reference
axis and $\dot \theta_i$ the corresponding conjugate variable. The potential conventionally
assumed is
\begin{equation}
V(x)=A[1-\cos(\omega x)],
\end{equation}
where, in order to keep the maximum force between any two neighboring rotors a constant
(set to unity) independent of the two parameters $A$ and $\omega$, they are set to satisfy
$\omega=1/A$, so that $A$ is adopted as the only independent parameter~\cite{SJ2020}.

In this homogeneous rotor lattice, a striking property is that there exists a transition
temperature, $T^{tr}$, governed by the interaction strength, $A$, below and above which the
heat conduction is of ballistic and diffusive type, respectively, characterized accordingly
by a divergent and convergent heat conductivity~\cite{RL-Livi, RL-Savin, RL-Spohn, RL-Dhar}.
Specifically, it has been established that $T^{tr} = A/5$ approximately~\cite{SJ2020}. As
such we should be able to build a thermal diode by introducing the graded interaction strength
to ensure that, when the heat current flows in the direction the interaction strength decreases,
the local temperature keeps below the local transition temperature throughout, so that the
current is strong. But when the current flows in the opposite direction, the local temperature
is above the local transition temperature at the high temperature end, so that this part
of the lattice plays an impeding role to resist the current. This idea has been verified to be
valid~\cite{SJ2020}. Specifically, for an interaction-strength graded rotor lattice, its
Hamiltonian is still given by Eq.~(1), but with the potential term $V$ being replace by the
local interaction potential
\begin{equation}
V_i (\theta_{i+1}-\theta_i)= A_i(1-\cos[\omega_i (\theta_{i+1}-\theta_i)]),
\end{equation}
where $\omega_i$ is fixed to be $\omega_i=1/A_i$ and $A_i$ specifies the local interaction
strength. The local transition temperature thus reads as $T^{tr}_i=A_i/5$.

Without loss of generality, suppose that $A_i$ changes from $A_0=A_L$ to $A_{N}=A_R$ with
$A_L$ and $A_R$ being two parameters satisfying $A_L > A_R$. Moreover, we refer to the direction
from left to right, i.e., from the first to the last rotor, the forward direction. For our aim
here, we restrict ourselves to investigate the graded lattices whose local interaction strength
is specified by the following cubic polynomial function:
\begin{equation}
g(x)=a_0 + a_1 x + a_2 x^2 + a_3 x^3,
\end{equation}
i.e., we set the rescaled position of the $i$th rotor as $x_i=i/N$, and then the corresponding
interaction strength is given as $A_i=g(x_i)$. The four coefficients $a_0, \cdots, a_3$ are set
by the ``boundary" conditions that, at the two ends, $g(0)=A_L$ and $g(1)=A_R$, and additionally,
$g'(0)=k_L$ and $g'(1)=k_R$, with the tangent $k_L$ and $k_R$ at the two ends respectively being
two additional parameters for us to control the local interaction strength in between. The special
case of the linearly graded lattice studied previously~\cite{SJ2020} corresponds to $a_2=a_3=0$,
or equivalently, $k_L=k_R=A_R-A_L$, and any other case can be seen as a perturbation to this
special one. With the freedom to assign $k_L$ and $k_R$, in the following we will scrutinize
how the TR efficiency would change as $k_L$ and $k_R$ for a given pair of values of $A_L$ and
$A_R$ with which the linearly graded lattice functions as a thermal diode. Surely, by considering
polynomial functions of high order or other forms of functions would allow us to study more
complicated and comprehensive perturbations; this will be discussed later.

\begin{figure*}[!t]
\hskip-1.0cm
\includegraphics[width=16.7cm]{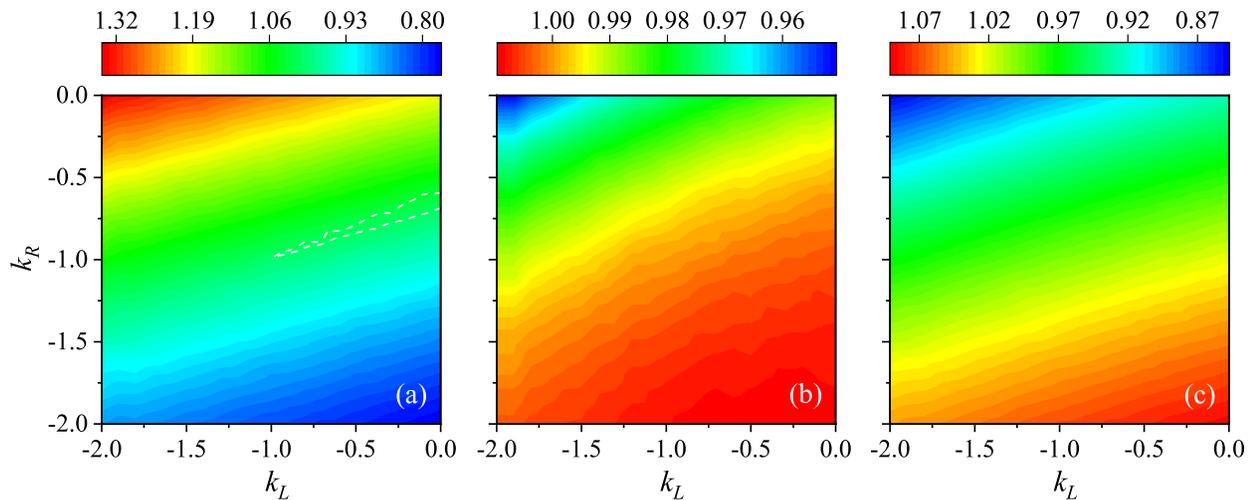}
\caption{The contour plots for the rescaled TR efficiency $\tilde{\mathcal{E}}$ (a), forward
current $\tilde{j_f}$ (b), and reverse current $\tilde{j_r}$ (c) in the interaction-strength
graded rotor lattice of size $N=512$. The white dashed lines in (a) are for $\tilde{j_f}=1$
(above) and $\tilde{j_r}=1$ (below), respectively. For a given pair of values of $k_L$ and
$k_R$, the local interaction strength is assigned with a certain cubic polynomial function
(see text).}
\end{figure*}

Next, to measure the TR efficiency, we perform the molecular dynamics simulations. To this
end, the system is coupled to two extra rotor lattice segments of $N_L$ and $N_R$ rotors
(sizes), respectively, at its left and right sides. Besides the neighboring interactions
the same as in the system but with homogeneous strength of $A_L$ and $A_R$, the motion of
the rotors in these two segments are also subject to the applied Langevin heat baths of
temperature $T_L$ and $T_R$, so that their motion equations are
$\ddot{\theta}_k = -\frac{\partial}{\partial \theta_k}{[V_k(\theta_{k+1}-\theta_k)+V_{k-1}
(\theta_{k}-\theta_{k-1})]}-\gamma \dot{\theta}_k+\xi_k$ with $\xi_k$ being a white Gaussian
noise satisfying that $\langle\xi_k(t)\xi_k(t') \rangle=2\gamma k_B T_{L,R} \delta(t-t')$.
Here $\gamma$ governs the coupling strength between the rotor and the heat bath and $k_B$ is
the Boltzmann constant (set to unity). For the system rotors in between these two segments,
their motion equations are $\ddot{\theta}_i=-\frac{\partial}{\partial \theta_i} {[V_i
(\theta_{i+1}-\theta_i)+V_{i-1}(\theta_{i}-\theta_{i-1})]}$. With all the motion equations,
the whole system is integrated numerically with a standard algorithm. When the system has
relaxed into the stationary state, the heat current $j$ is measured as the time average of
the local current $j_i$, i.e., $j=\langle j_i \rangle$, where $j_i$ can be defined as
$j_i=-\dot \theta_i \frac{\partial}{\partial \theta_i} V_i (\theta_{i+1} -\theta_i)$~\cite{
Lepri2003}. To obtain the TR efficiency at a given working temperature $T$ with a given bias
$\Delta T > 0$, let us denote the high and low boundary temperatures by $T_+=T+\Delta T/2$
and $T_-=T-\Delta T/2$, respectively; the forward current $j_f$ is thus measured numerically
by setting $T_L=T_+$ and $T_R=T_-$ and the reverse current $j_r$ by  $T_R=T_+$ and $T_L=T_-$.
Then the TR efficiency can be measured as~\cite{Walker}
\begin{equation}
\mathcal{E}=\frac{j_f-j_r}{j_f+j_r}.
\end{equation}
Here both $j_f$ and $j_r$ represent their absolute values. In our simulations, we have
adopted the velocity-Verlet algorithm~\cite{VV} and set $N_L=N_R=16$ and $\gamma=1$, but it
has been checked and verified that the results do not depend on these particular adoptions.

\section{Results}

Now we are ready to present the simulation results. For our aim here, in the following we
will focus on the typical case where $A_L=1.5$, $A_R=0.5$, $T=0.15$, and $\Delta T=0.2$.
The linearly graded lattice with this set of parameter values has been found to be an ideal
thermal diode whose TR efficiency keeps growing as the system size~\cite{SJ2020}. With the
setup detailed above, our particular interest is to reveal how the TR efficiency $\mathcal{E}$
would change when $k_L$ and $k_R$ deviate from $k_L=k_R=A_R-A_L$ of the linearly graded lattice.

Our main results are summarized in Fig.~2, where the TR efficiency, the forward and reverse
current, are presented as functions of $k_L$ and $k_R$, respectively. For the sake of
comparison, shown in Fig.~2 are the values rescaled by those of the linearly graded lattice,
$\mathcal{E}^{lin}$, $j^{lin}_f$, and $j^{lin}_r$, respectively; i.e., $\tilde{\mathcal{E}}=
\mathcal{E}/\mathcal{E}^{lin}$, $\tilde{j_f}=j_f/j^{lin}_f$, and $\tilde{j_r}=j_r/j^{lin}_r$.
Note that in all three panels of Fig.~2, the center point ($k_L=k_R=-1$) corresponds to the
linearly graded lattice. Above all, as Fig.~2(a) shows, the TR efficiency has a by no means
trivial dependence on $k_L$ and $k_R$. Over the investigated range $[-2,0]\times[-2,0]$ of
$k_L$ and $k_R$, $\mathcal{E}$ may undergo a considerable variation up to about $60\%$ of
$\mathcal{E}^{lin}$. In particular, an increase as high as over $30\%$ of $\mathcal{E}^{lin}$
can be reached (see for $k_L=-2$ and $k_R=0$), suggesting convincingly that the perturbation
effect is worth considering to enhance TR. On the other hand, the perturbation may lower
the TR efficiency as well. However, taking into account the information of the forward and 
the reverse current [see Figs.~2(b) and 2(c)], it may provide us with more flexibility for
designing the thermal diode of certain functions. For example, if we take the values of $k_L$
and $k_R$ at the bottom-right corner, then the resultant thermal diode would have such a
property that the forward current remains close to its maximum, whereas the TR efficiency
is dominantly determined by the reverse current that depends on $k_L$ and $k_R$ much more
sensitively. Of particular advantage is the area bounded by the two white dashed lines
[see Fig.~2(a)], where not only the TR efficiency is higher than the linearly graded lattice,
but also the forward current is stronger meanwhile the reverse current is weaker, respectively,
than their counterparts in the latter.

\begin{figure}[!]
\vskip-0.03cm
\includegraphics[width=8.0cm]{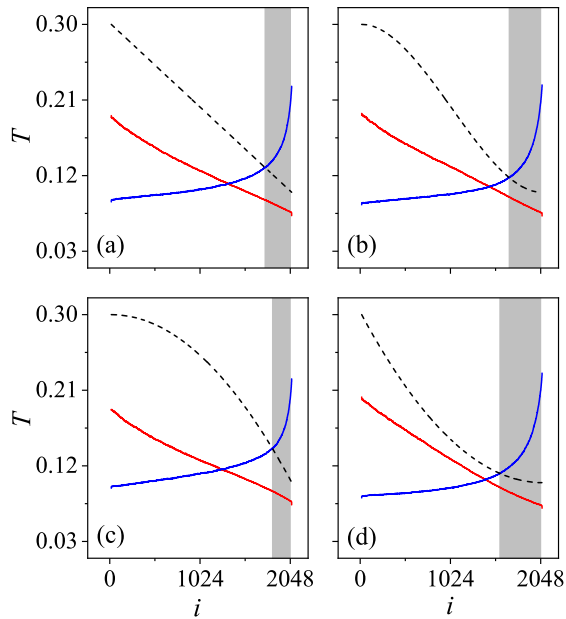}
\caption{The temperature profiles of the graded rotor lattice for (a) $k_L=k_R=-1$ (the
linearly graded lattice), (b) $k_L=k_R=0$, (c) $k_L=0$ and $k_R=-2$, and (d) $k_L=-2$ and
$k_R=0$. In each panel, the red (blue) line is for that when the thermal current flows
forwardly (reversely), the black dashed line is for the local transition temperature
$T^{tr}_i=A_i/5$, and the shaded area indicates the segment of the lattice
where the temperature is above the transition temperature when the current flows
reversely. The system size is $N=2048$.}
\end{figure}

In order to understand the perturbation effect emerges in Fig.~2, we study the temperature
profiles of four representative cases (see Fig.~3). First, even for the linearly
graded rotor lattice [see Fig.~3(a)], the temperature profile is obviously far from a
straight line when the current flows reversely, in spite of the small, constant gradient
of the local interaction strength, which is about $0.0005$ in this case.
The key role for forming such a curved temperature profile is played by the right end
segment (the shaded part) where the local temperature is higher than the local transition
temperature. As such the heat conduction is suppressed in this segment, making it an effective
thermal insulator. As a result, the temperature drops rapidly over this segment. On the other
hand, for the left segment, the local temperature is below the local transition temperature,
and hence it serves instead as a thermal conductor, over which the temperature drop must be
lower. This explains why the temperature profile for the reverse current is characterized,
respectively, by two segments of a mild and a rapid change. From this analysis we can
see that, as mentioned above, this strong nonlinear effect is indeed rooted in the
transition. As to the forward current, the local temperature is below the local transition
temperature throughout; therefore the whole lattice works as a thermal conductor, leading
to a strong forward current as well as a roughly linear temperature profile.

Importantly, as indicated by other panels of Fig.~3, the mechanism of transition
works generally even when the linearly graded structure is perturbed, showing that this
mechanism is quite robust. Comparing the temperature profiles of all four cases, it can
be seen that the difference between them is slight, implying that the system has a strong
adaptive ability to stabilize the temperature profiles against the perturbation. In fact,
the perturbations in the three perturbed cases are not very weak, which can be told directly
from the $T^{tr}_i$ curve that represents the local interaction strength $A_i$ as well
due to $T^{tr}_i=A_i/5$. Indeed, for the three perturbed cases, the deviation of $A_i$ from
the linearly graded rotor lattice is obvious. In addition, the robustness of the transition
mechanism also reflects in the fact that the TR behavior in the perturbed cases can be
explained based on the $T^{tr}_i$ and temperature curves. For example, comparing Figs.~3(c)
and 3(d), the temperature profile for the forward current (the red curve) lies much lower
below $T^{tr}_i$ in the former, suggesting that the forward current should be stronger in
the former than in the latter, which is in agreement with the results in Fig.~2(b). Similarly,
for the reverse current (the blue curve), as the resisting layer (shaded) is thicker in the
latter, a weaker reverse current is therefore expected in the latter, agreeing with the result
in Fig.~2(c). Due to these robustness features of the transition mechanism, it is interesting
to note that the TR behavior of a perturbed graded rotor lattice can be qualitatively
predicted without performing the simulations, because we can take the temperature profiles
of the linearly graded rotor lattice as approximations, and compare them with the transition
temperature curve $T^{tr}_i=A_i/5$ for the given interaction strength $A_i$.

\begin{figure}[!t]
\includegraphics[width=8.0cm]{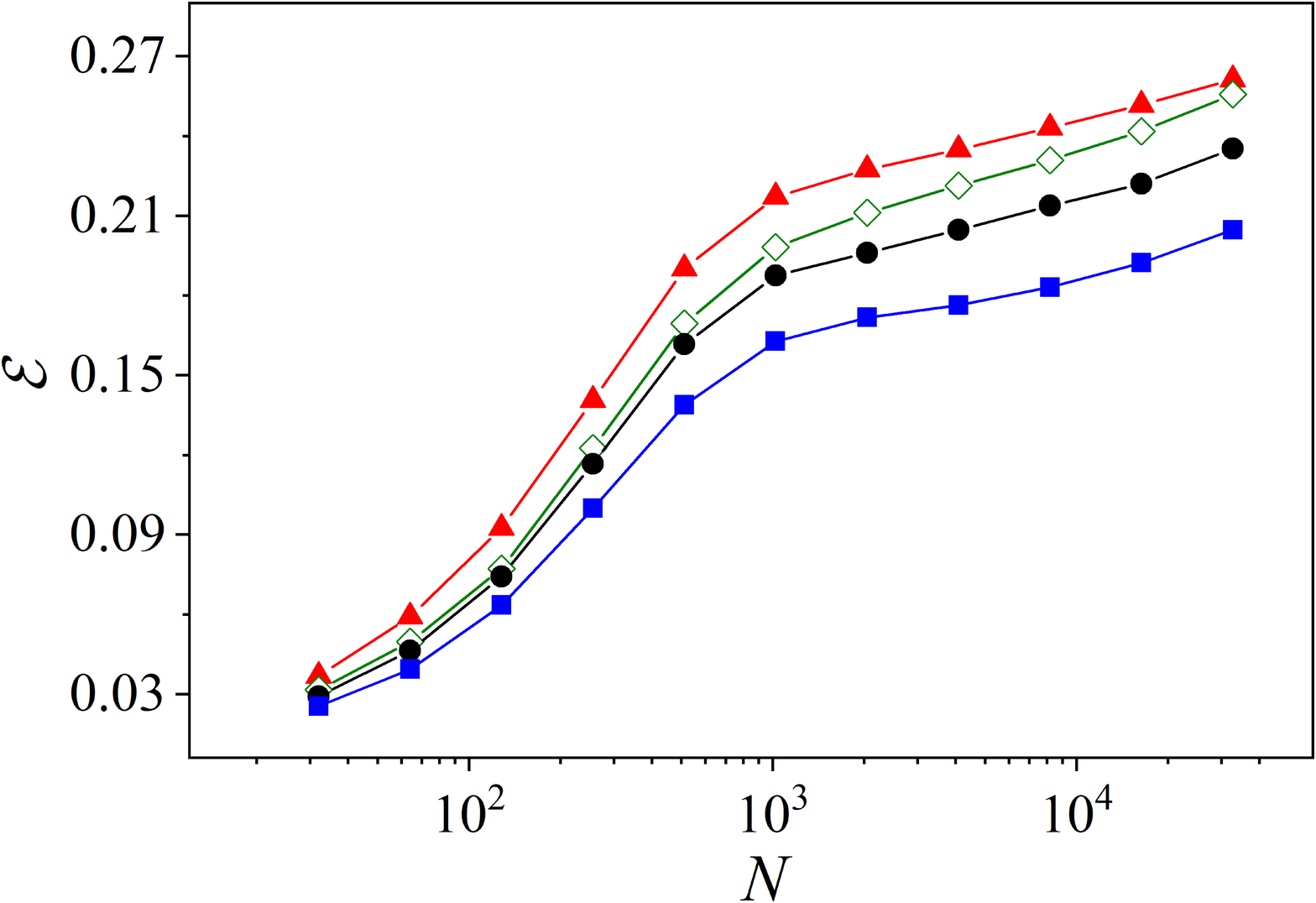}
\caption{The system size dependence of the TR efficiency of the graded rotor lattice for
several representative cases: $k_L=k_R=-1$ (black dots), the linearly graded case; $k_L=k_R=0$
(red triangles); $k_L=k_R=-2$ (blue squares); and $k_L=0$, $k_R=-0.486$
(green open diamonds).}
\end{figure}

As a crucial issue for our motivation here, we need to investigate if the robustness
features addressed above would survive the thermodynamical limit. Our simulation results
suggest a positive answer to this question. To this end, several representative cases are
simulated with various system sizes, and the results for the TR efficiency are presented
in Fig.~4. It can be seen that, accompanying with the curve for the linearly graded rotor
lattice, all other curves for the illustrative perturbed cases keep to grow together.
This property is welcomed; it shows that the perturbation effect is robust in the
thermodynamical limit and hence may be utilized to improve the TR efficiency for 
any size of the system.

\begin{figure}[!t]
\includegraphics[width=8.0cm]{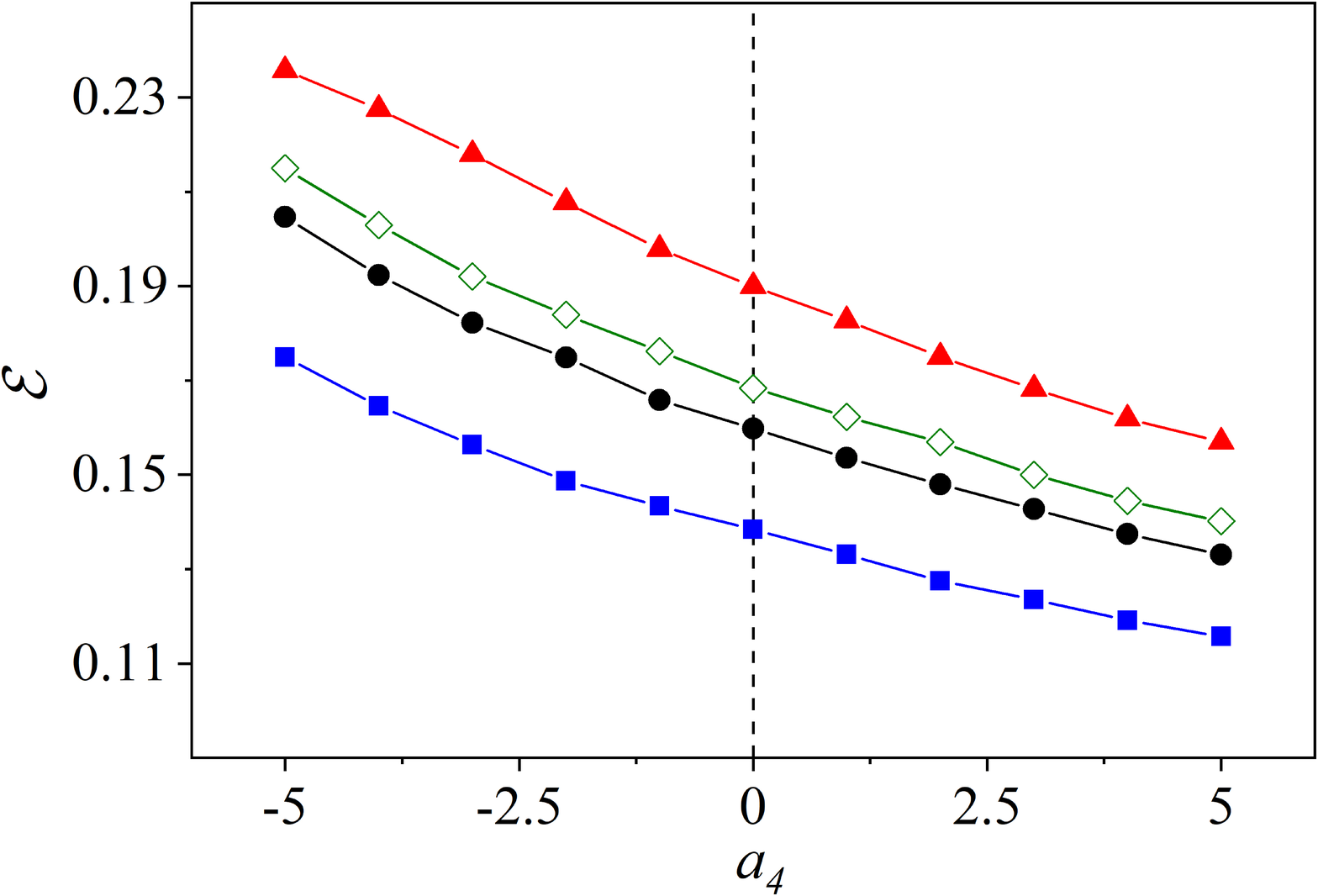}
\caption{The TR efficiency for the graded rotor lattice whose local interaction strength
is assigned with the fourth-order polynomial function where $a_4$ is the coefficient of
the highest term. The symbols on the dashed line indicates the TR efficiency for $a_4=0$,
corresponding to results when the cubic polynomial function is adopted. The system size
is $N=512$; $k_L=k_R=-1$ (black dots), the linearly graded case; $k_L=k_R=0$ (red triangles); 
$k_L=k_R=-2$ (blue squares); and $k_L=0$, $k_R=-0.486$ (green open diamonds).}
\end{figure}

So far we have focused on the polynomial perturbation up to the cubic term for two reasons. 
One is that we have supposed such a form may have captured the most significant part of a 
perturbation. Another is that, as such we have two free parameters ($k_L$ and $k_R$), the 
two-dimensional parameter space spanned by them has been big enough to explore in view of 
our computing resources, which have been fully engaged in carrying out a detailed investigation 
as in Fig.~2. But what effect a complicated perturbation may have is not clear yet, and it 
seems to be hard to predict in view of the strong nonlinear effect we have witnessed. Just 
as a preliminary attempt, we show in Fig.~5 how the TR efficiency may respond if the 
fourth-order term, $a_4 x^4$, is added to the function $g(x)$ given in Eq.~(4), where, 
for a given value of $a_4$, other four parameters $a_0, \cdots, a_3$ are determined in 
the same way. Undoubtedly, as shown in Fig.~5, the effect it induces could be significant, 
justifying that more complicated perturbations deserve further study.

Finally, as a comparison, in the following we conduct a parallel study of the mass graded
Fermi-Pasta-Ulam-Tsingou (FPUT) lattice~\cite{FPU-1, FPU-2}. Its Hamiltonian is
\begin{equation}
H=\sum_i [\frac{p_i^2}{2m_i}+V(x_{i+1}-x_i-1)],
\end{equation}
where $x_i$ and $p_i$ are the conjugate variable pair of the $i$th particle, $m_i$ its mass,
and $V(x)=x^2/2+x^4/4$. The mass graded FPUT model was first come up with in
Ref.~\cite{YNPRB}, where the TR was illustrated for the first time in the graded structure.
Specifically, it is reported in Ref.~\cite{YNPRB} that when the masses of the $N$ particles
of the system are assigned to change linearly from $m_1=m_L=10$ to $m_N=m_R=1$, a relatively
stronger TR effect would be observed at the working temperature $T=0.1$ with the system size
$N=200$.

\begin{figure*}[!]
\includegraphics[width=17.7cm]{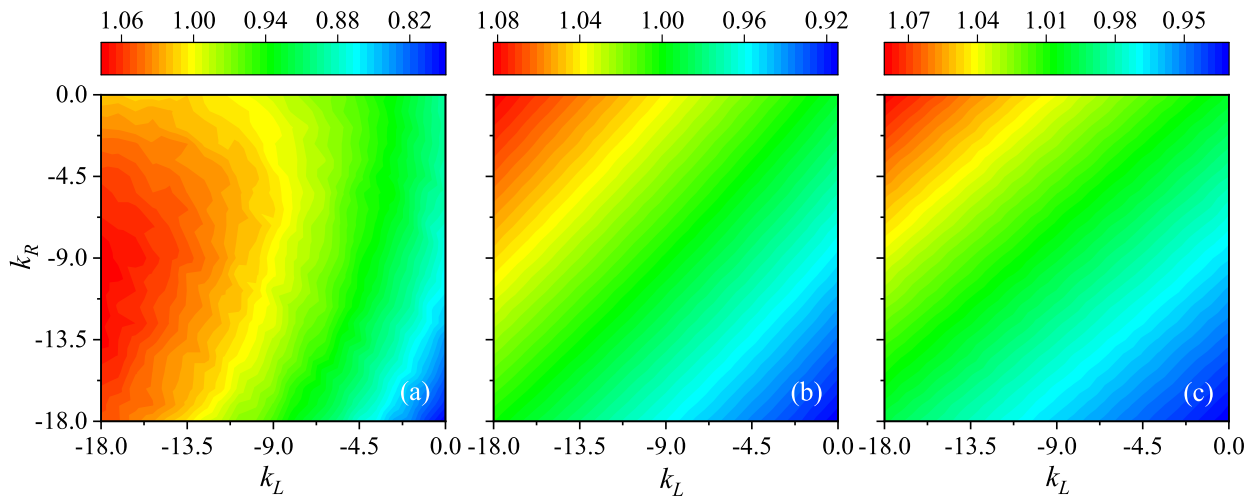}
\caption{The contour plots for the rescaled TR efficiency $\tilde{\mathcal{E}}$ (a), forward
current $\tilde{j_f}$ (b), and reverse current $\tilde{j_r}$ (c) for the mass graded FPUT
lattice. The system size, the working temperature and the temperature bias are, respectively,
$N=200$, $T=0.1$, and $\Delta T=0.1$. See text for the masses of particles assigned for each
pair of $k_L$ and $k_R$ values.}
\end{figure*}

As our simulation results suggest (see in the following), with the
increasing size, the TR effect would decay till disappear. This is in clear contrast with
the interaction-strength graded rotor lattice. Our motivation here is to see what the
perturbation to the linearly graded structure may lead to in this category of TR systems.
Similarly, in this case we set the graded masses of particles according to the cubic
polynomial function of Eq.~(4) as well; i.e., $m_i=g(\bar x_i)$, where $\bar x_i=i/N$ is
the rescaled average position of the $i$th particle. The four prefactors in $g(x)$ are set
following the conditions that $g(0)=m_L=10$, $g(1)=m_R=1$, $g'(0)=k_L$, and $g'(1)=k_R$.
As such the case of linearly graded masses corresponds to $k_L=k_R=-9$, and we take it
as the reference again. The working temperature and temperature bias are set to be
$T=\Delta T=0.1$. The simulation method and the related parameter values are the same
as adopted in the simulations of the rotor lattice.

First, a counterpart of Fig.~2 but here for the mass graded FPUT lattice is presented as
Fig.~6 for $N=200$ that ensures roughly the strongest TR efficiency. In Fig.~6(a), it can
be recognized that, though more moderate than in the graded rotor lattice, the perturbation
can still result in a significant change in the TR efficiency, from about $20\%$ below to
$10\%$ above $\mathcal{E}^{lin}$ (corresponding to the center point), over the square range
of $k_L$ and $k_R$ investigated. Therefore, even for this category of TR systems, there
is still space for improving the TR effect by adapting the nonlinearly graded structure.
On the other hand, such a study as in Fig.~6 may also help for revealing more features of
the studied object. Taking the current case as an example, it shows [see Fig.~6(a)] that
generally the TR efficiency has a more sensitive dependence on $k_L$ rather than $k_R$,
suggesting that the detailed mass distribution over the heavy end of the system could matter
more.

Next, let us probe how the TR efficiency may vary versus the system size. The results for
three representative cases are illustrated in Fig.~7. It is interesting to note that, as
in the linearly mass graded case, in the perturbed cases the TR efficiency hits its maximum
more or less around the size of $N=200$ as well. Moreover, the variant range or $\mathcal{E}$
due to the perturbation is also the widest at the same size. Decreasing or increasing the system
size, neither the TR nor the perturbation effect on TR would be sustained, indicating that 
neither is robust against the change of the system size.

\section{Discussion and conclusions}

\begin{figure}[!]
\includegraphics[width=8.0cm]{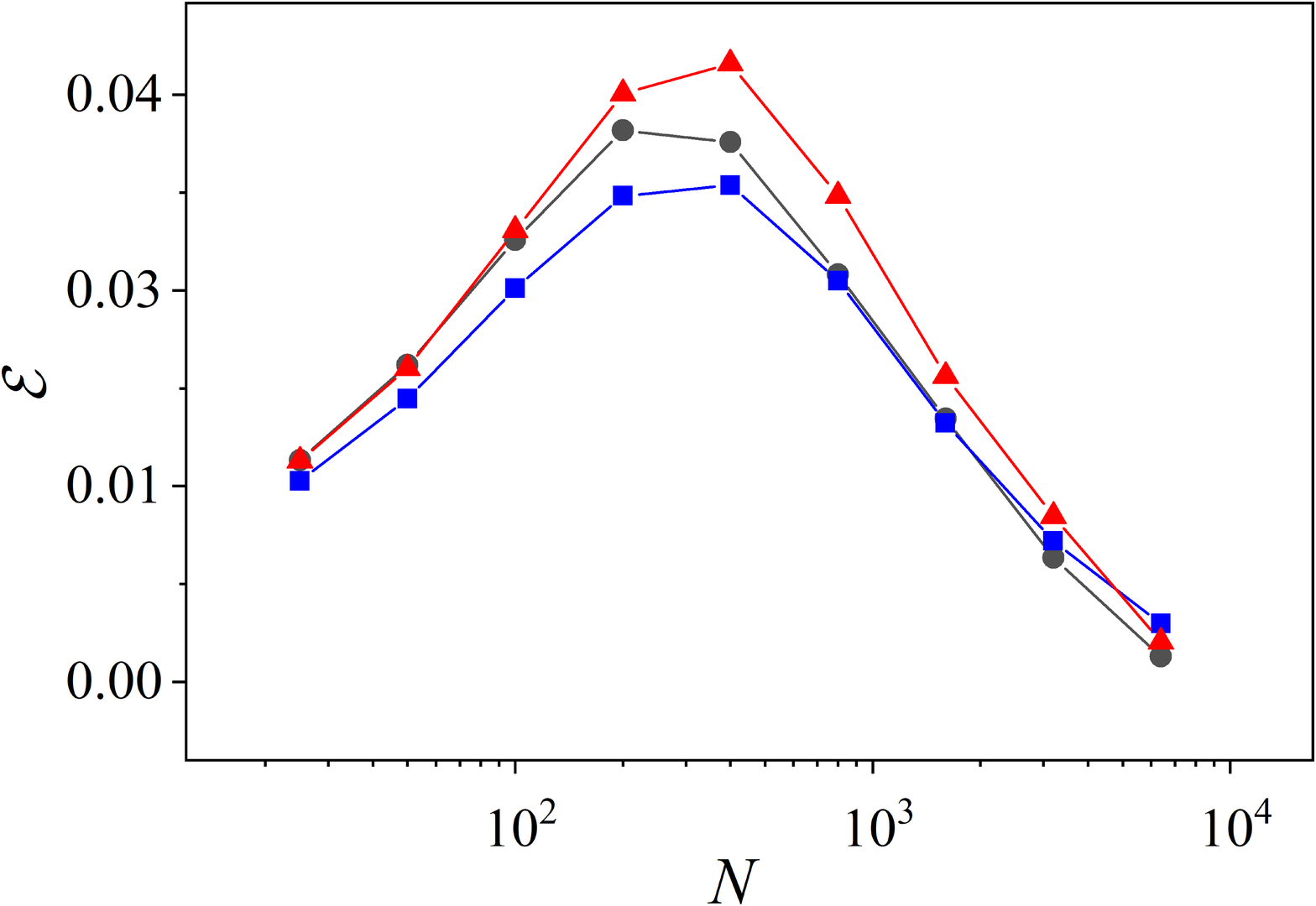}
\caption{The system size dependence of the TR efficiency of the mass graded FPUT lattices.
The black squares are for the linearly graded case with $k_L=k_R=-9$; the red triangles and
blue squares are for, respectively, the graded masses determined by $k_L= - 18, k_R = -9$
and $k_L=k_R=0$.}
\end{figure}

In summary, we have investigated how the graded structure may influence TR with two paradigmatic
lattice models as representatives of two distinct categories. We have mainly
focused on the possible effect a perturbation to the linearly graded structure may induce. In
terms of the TR efficiency, it has been found that, the perturbation effect depends on the TR
efficiency of the linearly graded structure itself: The larger the TR efficiency in the latter,
the larger the variation to it the perturbation may bring in. Therefore, for a good thermal diode,
adaption of the graded structure may serve as an effective tactic to optimize its TR efficiency.
Besides, the study from the perturbation perspective has also been found to be effective to reveal
other interesting properties, such as the strong adaptive ability to maintain the temperature
profiles in the graded rotor lattice.

In the present study, only the graded structures described by cubic polynomial functions are
concerned. Though other graded structures are worth studying, the challenge of simulation has
to be faced. In this respect, the machine learning technique may provide a powerful tool.
Indeed, as illustrated in a series of recent studies, the machine learning method has been
found superior in studies of thermal conduction issue (see Ref.~\cite{ML} for a recent review
and references cited therein). Its advantage lies in that the brutal
searching in the parameter space can be avoided; instead, aiming at the prescribed TR target,
the learning machine may figure out the shortcuts to the target. For the TR effect that sustains
in the thermodynamical limit, the structure of the known models, such as that with the integrable
spacer and the graded rotor lattice, can be used as input for training. It would be rewarding
if new mechanisms and models unknown yet can be solved out by this strategy.

This work is supported by the National Natural Science Foundation of China (Grants No. 12075198
and No. 12047501).



\end{document}